\documentclass[aps,pre,superscriptaddress, twocolumn]{revtex4-1}

\usepackage[dvips]{graphicx}
\usepackage{amsmath}
\usepackage{amssymb}
\usepackage{natbib}
\usepackage{psfrag}
\usepackage{color}
\usepackage{epsfig}
\usepackage{subfigure}
\begin{document}

\title{Hydrodynamically synchronized states in active colloidal  arrays}

\author{{Lo{\"{\i}}c} Damet}
\affiliation{Cavendish Laboratory and
  Nanoscience Centre, University of Cambridge, Cambridge CB3 0HE,
  U.~K.}

\author{Giovanni M. Cicuta}
\affiliation{Dip. Fisica, Universit\`{a} di Parma, and INFN, Sez. Milano-Bicocca, gruppo di Parma, Italy}

\author{Jurij Kotar}
\affiliation{Cavendish Laboratory and
  Nanoscience Centre, University of Cambridge, Cambridge CB3 0HE,
  U.~K.}

\author{Marco Cosentino Lagomarsino}
\affiliation{Genomic Physics Group, FRE 3214 CNRS ``Microorganism Genomics'' and University Pierre et Marie Curie, 15, rue de l'\'{E}cole de M\'{e}decine Paris, France}

\author{Pietro Cicuta}
\affiliation{Cavendish Laboratory and
  Nanoscience Centre, University of Cambridge, Cambridge CB3 0HE,
  U.~K.} \email[correspondence to:]{pc245@cam.ac.uk}


\begin{abstract}
 Colloidal particles moving in a fluid interact via the induced velocity field.  The collective dynamic  state for a class of actively forced  colloids, driven by harmonic potentials via a rule that couples forces to configurations, to perform small oscillations around an average position, is shown by experiment, simulation and theoretical arguments to be determined by the eigenmode structure of the coupling matrix. It is remarkable that the dynamical state can therefore be predicted from the mean spatial configuration of the active colloids, or from an analysis of the fluctuations near equilibrium. This has the surprising consequence that while  2 particles, or polygonal arrays of 4 or more colloids,  synchronize with the nearest neighbors in anti-phase, a system of 3 equally spaced colloids synchronizes in-phase. In the absence of thermal fluctuations, the stable dynamical state is predominantly formed by the eigenmode with longest relaxation time.   
\end{abstract}


\maketitle

\begin{figure*}[t]
\begin{center}
\epsfig{file=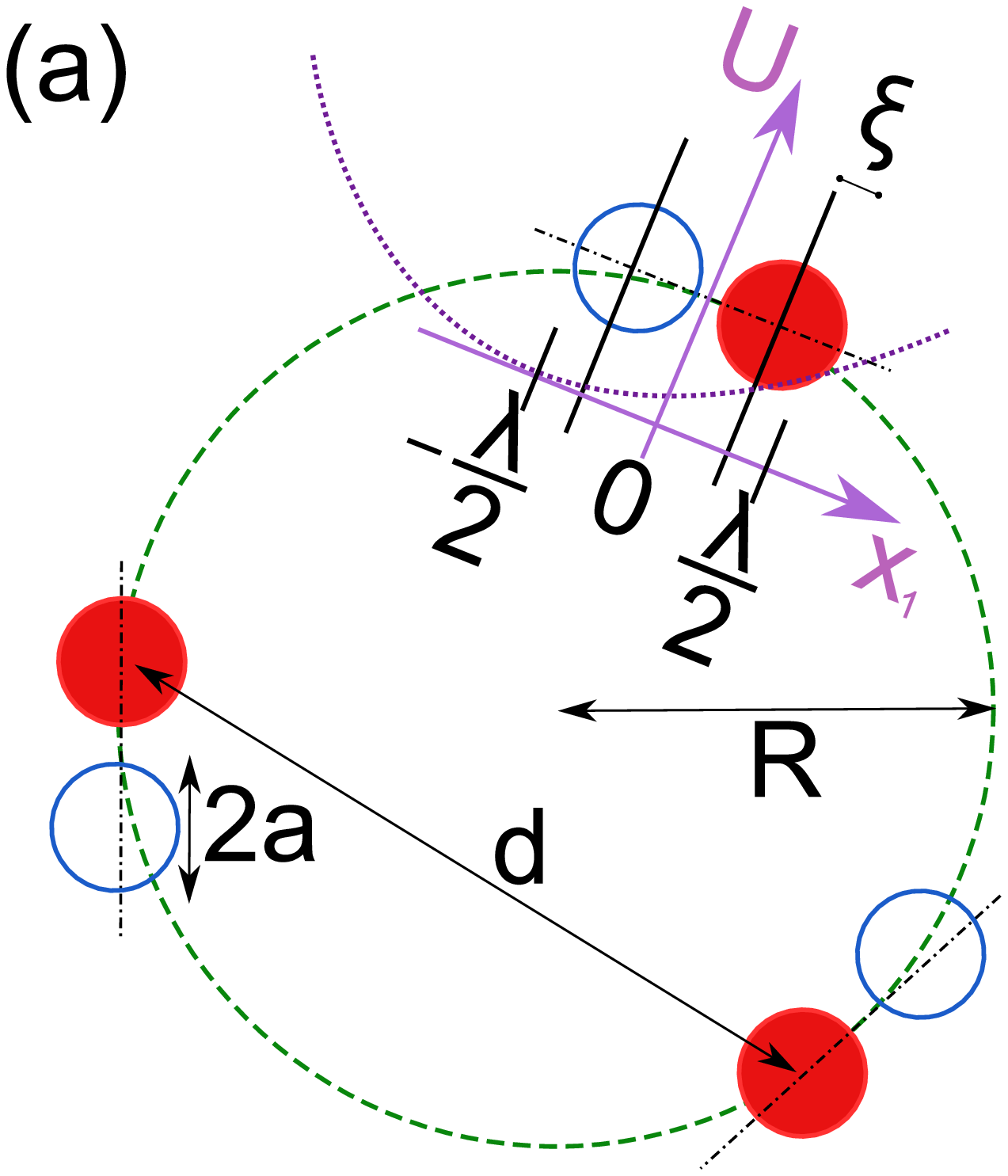, width=3.2cm}\,\,\,
\epsfig{file=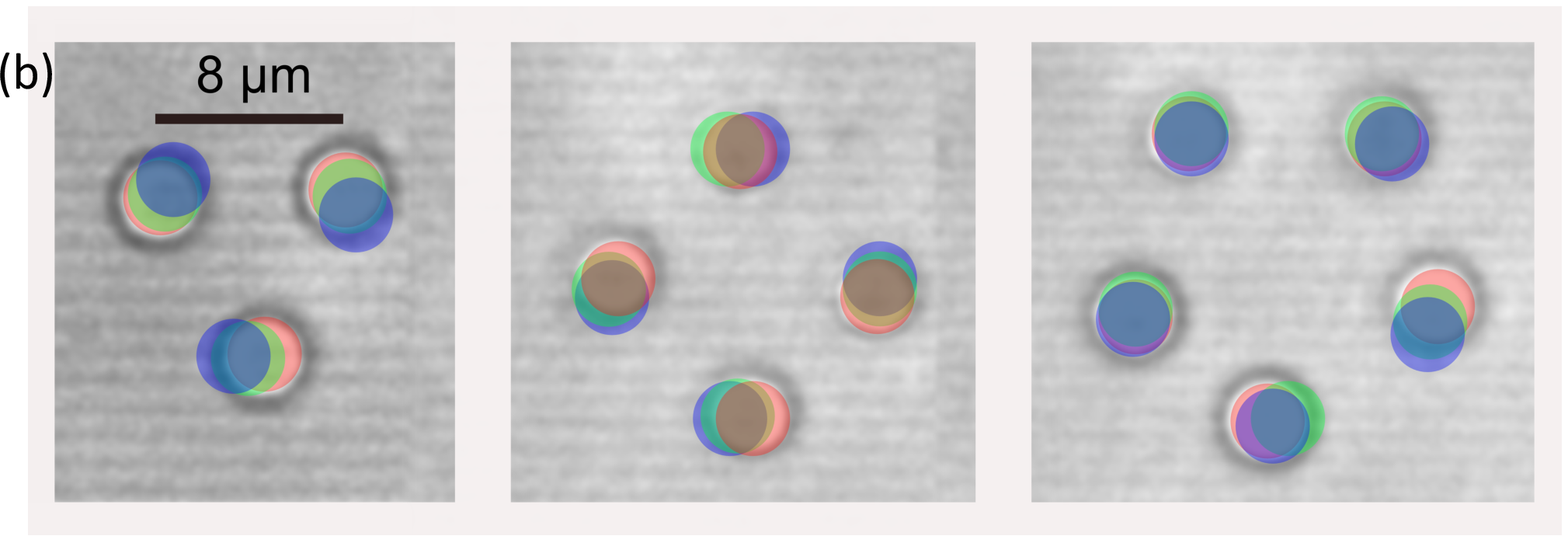, width=12cm}
\caption{(Color online) \textbf{Regular arrays of actively driven  colloidal particles synchronize into   steady collective dynamical states. }(a) sketch of experiment. As the number of particles $N$ is varied from $N=3$ to $N=5$ (and beyond, numerically), the   circle radius    $R$ is increased to maintain constant
 the distance between bead centers  $d=2R\sin(\pi/N)=8\,\mu$m.    (b) Images showing one snapshot of the system, where the particles are highlighted in red.  Particle positions after 20 frames (green) and 40 frames (blue) are overlayed. Videos are available as SM, and show clearly the $N=3$ system performing in-phase oscillations, the $N=4$ with neighbors in anti-phase, and $N=5$ with phase locking between neighbors. The optical tweezers system performs image analysis on every frame to determine particle positions, and thus implement the ``geometric switch'' condition described in the text, at  rates between 200 and 300 frames/s. Each colloidal particle is effectively a phase oscillator, undergoing a motion bound in amplitude but free in period and phase.
   \label{fig1}}
\end{center}
\end{figure*}


Hydrodynamic coupling at low Reynolds' number is an important feature in biological flows, with  key consequences in apparently diverse phenomena such as the motility of microorganisms~\cite{goldstein09a}, circulation in the brain~\cite{rakic10} and functioning of the ear~\cite{hudspeth11}.  In various biological tissues, a macroscopic number of  units, for example cilia, display synchronized dynamics, leading to metachronal waves~\cite{gueron97}. Nearby cilia may beat in-phase or out of phase, and indeed may be in a condition where is it possible to readily switch between the two dynamical states~\cite{goldstein09a}. One outstanding question is what determines the character of the dynamical steady state.
Recent progress in ``hydrodynamic synchronization'' is reviewed in~\cite{golestanian11b}, and an overview of  low Reynolds number (Re) flows~\cite{lauga09}. In an attempt to model (both experimentally and theoretically) the physics of hydrodynamic synchronization, two main ideas have emerged. The first is to consider the coupling of two or more objects driven by a constant force over general (pre-defined) orbits~\cite{lenz08}.  Within this idea of  ``rotor'' cilia models, the   phase of each rotor is free and there can be synchronization of different rotors under certain conditions. This  has recently been studied very generally in~\cite{golestanian11a}, extending the case of circular driving force orbits~\cite{lenz08}.  A different model consists of a ``geometric switch'', and was proposed in~\cite{bassetti03}; here the force is discontinuous, and this is not described within the formalism of~\cite{golestanian11a}. It is not yet established which of these ideas is most appropriate to describe a biological scenario. The discontinuity of force should not be discounted a-priori, considering the fact that molecular motors undergo discrete attachment/detachment events which couple to the force generation.  In previous work the geometric switch model was investigated for the simple case of  two active elements, showing the robustness of hydrodynamic coupling in the presence of noise~\cite{cicuta10a}. A linear chain of oscillators, following the geometric switch rule with general actuation forces, was studied recently by numerical methods, in the absence of noise~\cite{stark11}.

In this Letter we show how the non-equilibrium dynamical behavior of the ``geometric switch'' model can be understood from the eigenmode structure of the Oseen tensor, which depends  on the geometrical arrangement oscillators. Systems of a small number of elements are considered, in the presence of thermal noise: between three and five colloidal particles are arranged on equally spaced average positions on a circumference, and each one is maintained in a driven oscillatory tangential orbit, with fixed amplitude but free phase and period.    The main result  is  that the fundamental (longest lived) hydrodynamic mode (an easily derivable quantity, in contrast to finding full solutions of the system) dominates the collective motion in the driven steady state. As a consequence, the steady state  can be strikingly  different depending on the number of particles and their arrangement. The ``dynamical motifs'' observed here may guide the analysis of  cilia coordination in complex biological systems.

 This system is realized experimentally with optical traps, with fast video feedback to impose the fixed amplitude driven oscillation. The only interaction between the elements is through the  hydrodynamic flow arising from the colloid movements. Brownian Dynamics (BD) simulations, in which the hydrodynamic interaction is calculated through Oseen's tensor~\cite{ermak1978}, are compared to the experimental data and  can also be performed readily with a larger number of particles in the system.
 The assumptions for this treatment~\cite{brenner83} are a low Reynolds number, particles far relative to their diameter, and a steady flow, both of which are satisfied in the physical context.
 As a further consequence of hydrodynamic interaction, there are correlations in the  Brownian fluctuations of different particles.
  The dynamics of systems where  spheres are held by fixed harmonic potentials on the vertices of arbitrary planar regular polygons has been solved \emph{exactly} within  Oseen's description of hydrodynamics~\cite{cicuta10c}, giving a basis from which to understand the coupling in the active scenario.

\begin{figure}[t!]
\begin{center}
\epsfig{file=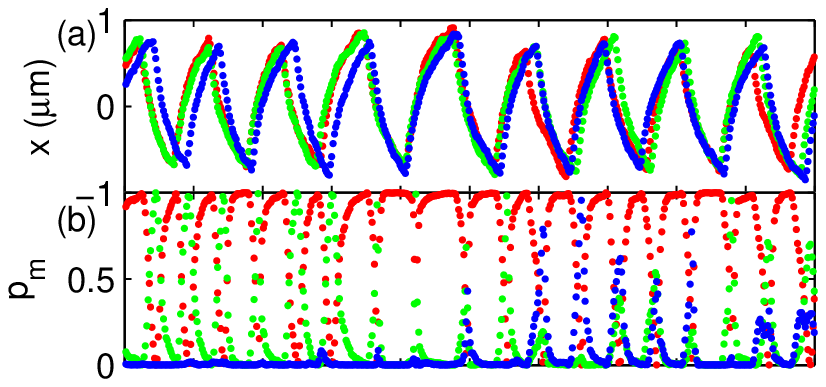, width=7cm}\\
\epsfig{file=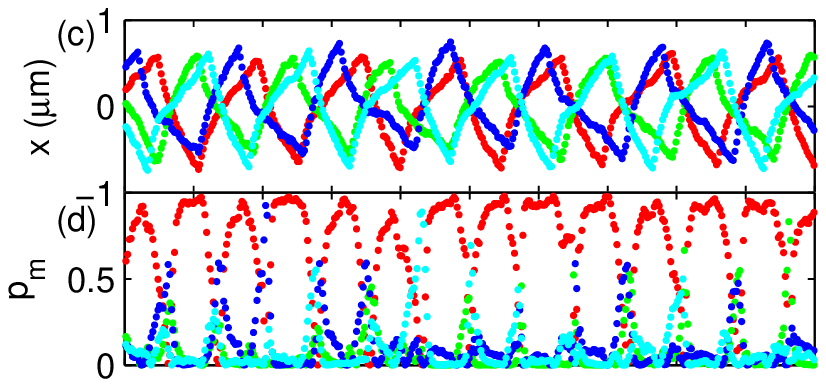, width=7cm}\\
\epsfig{file=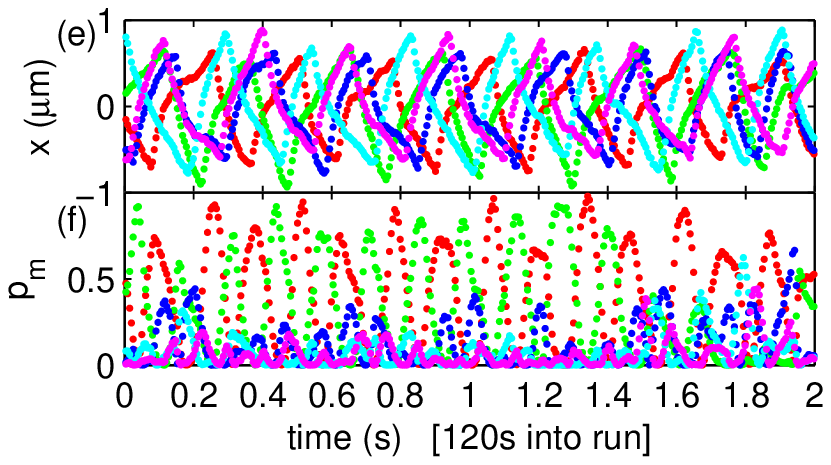, width=7cm}\\
\caption{(Color online) \textbf{The active tangential motion of beads is strongly correlated in experiments. } Shown here
are the displacements $x(t)$ relative to the mean position, for systems of $N=3,4,5$ in (a), (c), (e) respectively. The color code
red, green, blue, cyan, magenta identifies different beads, anticlockwise.  In (b), (c), (d) the instantaneous projections $p_m(t)$ $m=1,2,...N$ on the eigenvectors of the
coupling matrix show the important characteristic that the mode with longest relaxation time (red) dominates the displacement configuration. For $N=5$, as for larger odd-numbered systems, there are two degenerate modes (red, green), which are seen to alternate in amplitude.  The color code indicates decreasing relaxation time: red, green, blue, cyan, magenta.   \label{fig2}}
\end{center}
\end{figure}

 Optical traps are used to confine colloidal beads within harmonic
potentials, the system hardware is  described  in greater detail in ~\cite{cicuta09a,cicuta10a}. In this work, a varying number of silica beads of radius  $a=1.5\,\mu$m (Bangslabs) are  trapped from  below by a time-shared laser beam, focussed by a water immersion objective (Zeiss, Achroplan
IR  63x/0.90\,W). A pair of acousto-optical deflectors (AOD) allows the positioning of the laser beam in the $(x,y)$ plane with sub-nanometer precision; time-sharing is at a rate $\sim 10^5$\,Hz, corresponding to negligible diffusion of the beads in each laser cycle.
The solvent in which the beads are suspended is a  glycerol (Fisher, Analysis Grade) in water (Ultrapure grade, ELGA) solution 50\% w/w, giving a nominal  viscosity of $\eta=6$\,mPa\,s at 20$^o$C~\cite{CRCHandbook79}. Experiments are performed in a temperature controlled laboratory, $T=21^o$C.  The trapping plane is positioned $(20 \pm 1)\,\mu$m above the flat bottom of the sample, in a sample volume that is around 100\,$\mu$m thick.

  To realize the ``geometric switch'' condition, an active driving of each colloid is implemented here, similarly to~\cite{cicuta10a}, but for 3,4 or 5 particles, and driving the colloids on segments tangential to the ring on which they are positioned on average, see Figure~\ref{fig1}(a). A boundary is set at a pre-defined particle position, a distance $\xi$ from the minimum of the currently active optical trap.    The trapped particle moves (on average) towards the trap minimum and, when it  crosses the boundary, the current trap is switched off and the other trap, with its minimum a distance $\lambda$ away, is activated. Therefore the amplitude of oscillations is $\lambda-2\xi$.  This process is implemented via image analysis and feedback to the AOD for laser deflection in the experiments (and as a condition in the numerical simulations). Colloidal particles are always being driven, and  out of equilibrium in the sense that they do not reach the minimum of the active trap.  

  The optical trap potential is harmonic to a very good approximation, with stiffness $\kappa$ in the range  1.0 to 2.6\,pN/$\mu$m, depending on the number of beads trapped (each time, calibrated with precision $\pm$0.2\,pN/$\mu$m). The  relaxation time $\tau_0=\gamma/\kappa$ is of the order of 0.1s.
    The experiments have been performed with $\lambda=2\,\mu$m, $\xi= 0.31\,\mu$m and $d=8\,\mu$m.  The period of an isolated oscillator is $T_0=2\tau_0 \log[(\lambda-\xi)/\xi]$~\cite{cicuta10a}, which under the experimental conditions is about 0.3s. With image acquisition  through an AVT Marlin F-131B CMOS camera, operating at  shutter aperture time of  1.5\,ms, and frame rate  between 200 and 300\,fps (depending on the ring size, hence captured region of interest) there are multiple frames captured within the relevant timescales $\tau_0, T_0$.
  Since the configuration is analysed experimentally only at each frame, the corresponding time interval should be considered as a feedback time. Video is acquired  for over 4 minutes, i.e. over 48000\,frames. There is typically a transient lasting around a few periods before the systems reach the steady state discussed below.
In addition to the driven motion, colloids are affected by  stochastic thermal fluctuations fluctuations and by the net flow induced by all other moving particles.  Experiments are performed increasing the number of beads $N$, maintaining constant the arc-distance between neighboring beads as shown in Fig~\ref{fig1}(b).

\begin{figure}[t!]
\begin{center}
\epsfig{file=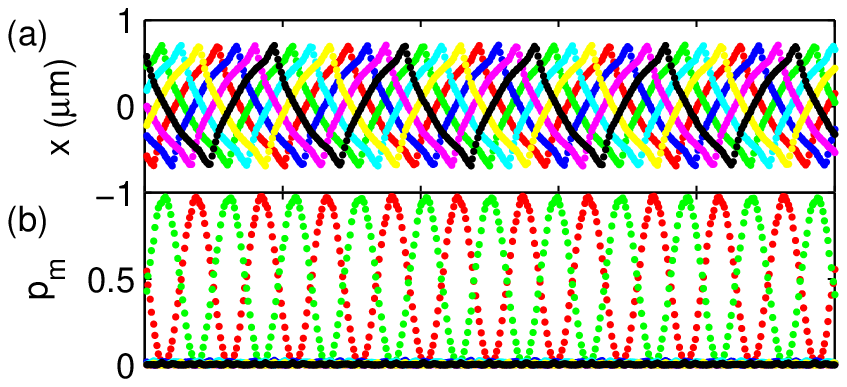, width=7cm}\\
\epsfig{file=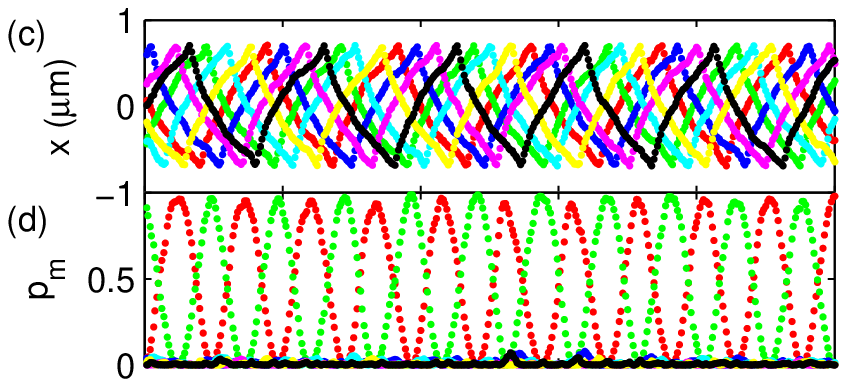, width=7cm}\\
\epsfig{file=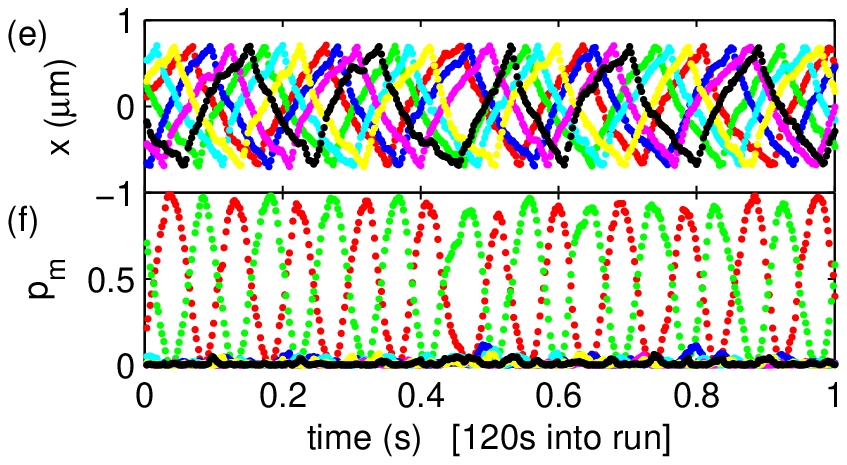, width=7cm}\\
\caption{(Color online) \textbf{The dynamic state converges to the analytical prediction in the limit of no Brownian noise.} In numerical simulation
it is straightforward to tune the level of noise.  Here, the trajectories and mode projections  of  a system of 7 particles are shown for temperatures of $T$=1K (a,b), 273K (c,d) and 1000K (e,f), with other physical parameters close to the experiments and $\kappa$=2\,pN/$\mu$m, $\xi=0.3\,\mu$m.  Colors match those in  Figures~\ref{fig2}, plus yellow and black used for beads (and modes) 6 and 7 respectively.     \label{fig4}}
\end{center}
\end{figure}



 Figure~\ref{fig2}(a) shows that for $N=3$ the three beads move in phase with each other, whilst for $N=4$ (Fig.~\ref{fig2}(c)) the nearest neighbors are in  anti-phase.  The behavior of $N=5$  (Fig.~\ref{fig2}(e)) is an apparently more complex phase locking. To understand the character of the steady state solution we explored the properties of the Oseen coupling tensor, which depends on the geometry of the mean particle arrangement.    Taking a step back, we recall that in passive systems the theoretical description of the fluctuations  of many-bead ring systems held in fixed position optical traps was calculated recently~\cite{cicuta10c}, extending original ideas by  Polin,  Grier and  Quake~\cite{quake99,quake06} and \cite{ruocco07}. A system of $N$ beads held in static traps in two dimensions has $2N$ normal modes, and these can be calculated analytically for configurations with high symmetry~\cite{quake06,cicuta10c}. The active driving of each bead along its tangent direction can be thought as introducing a constraint for each bead to move only along its fixed tangential direction. This reduces the number of modes by half, so that there are $N$ normal modes for a $N$ bead system. They can be obtained by projecting the two dimensional system of equations of motion~\cite{cicuta10c} onto the tangent vectors. Similarly to previous analysis of passive geometries~\cite{quake06,ruocco07,cicuta10c}
   the motion of the $j^{th}\mbox{-}$particle originates a force on the $i-$particle ${\vec f}_\alpha^{i,j}=(H^{-1})^{\alpha,\beta}_{i,j}v^{(j)}_\beta$ that depends on the whole configuration, where $H$ is the Oseen tensor~\cite{brenner83}. This leads  to a system of equations:
 \begin{eqnarray}
  \left\{
\begin{array}{ccc}
&{\vec F}_i-\sum_{j=1}^n H^{-1}_{i,j} \frac{d {\vec r}_j(t)}{dt}+{\vec f}_i(t) =0  , \,\, i=1,2\cdots , n \qquad\\
   &{\vec r}_i(t)\cdot {\vec t}({\vec r}_i)=0   , \,\, {\vec t}(\theta)=\left(
   \begin{array}{cc}
   -\sin \theta \\ \cos \theta \end{array}\right),
   \end{array} \right.
\label{aa.1}
\end{eqnarray}
in which the coupling force scales as $a/d$, and the  force ${\vec F}_i$ acting on the $i^{th}\mbox{-}$particle is  harmonic and tangent to the ring, with the ``geometric switch'' rule,
\begin{eqnarray}
{\vec F}_i=-\kappa\left[x_i(t)\mp \frac{\lambda}{2}\right]{\vec t}\left(\frac{2\pi (i-1)}{n}\right),
\label{aa.2}
\end{eqnarray}
where $\pm \frac{\lambda}{2}$ is the coordinate of the bottom of the harmonic well.  The   stochastic force $f_i(t)$ in Eq.~\ref{aa.1} represents the thermal noise on the $i^{th}\mbox{-}$particle, and it can be assumed that
  $<f_i(t)>=0$, $<f_i(t_1)f_j(t_2)>=2 k_BT \, (H^{-1})_{ij}\, \delta (t_1-t_2)$~\cite{quake06}.
${\vec t}({\vec r}_i)$ is a versor tangent to the ring, at the position ${\vec r}_i$, with anti-clockwise direction. The system of eq.~\ref{aa.1} can be linearized for small displacements. For symmetric arrangements   the coupling matrix is particularly simple, and the eigenvectors can be obtained readily~\cite{cicuta11w}. The eigenvectors are the normal modes of the coupled system.

 Knowing analytically the  normal modes of the system constrained to tangential motion, it is possible to decompose the dynamical steady state into projections onto each of the modes, and look at the mode  evolution in between switch events.  Experimentally it is clear that for $N=3$ and $N=4$, there is a single  mode which has very high amplitude (Fig~\ref{fig2}(b,d)).  In contrast, for $N=5$ two modes have high amplitude, and they alternate periodically (Fig~\ref{fig2}(f)).  This behavior is confirmed by BD simulations performed as in~\cite{cicuta10a}, see figure in Supplementary Materials.  
 
  Let us recall that
  for every \emph{even} value of $N$, the eigenmode  with longest relaxation time is described by the  pairs of adjacent spheres moving tangentially  in an anti-phase  motion, as shown in the companion paper~\cite{cicuta11w}.  For \emph{odd} $N$ with $N>3$, the longest relaxation time corresponds to two equivalent modes, in which the motion of neighbors is almost in anti-phase, but on cycling around the ring  each  particle is time-shifted by $+\Delta T$ (or  $-\Delta T$ in the equivalent mode) relative to its neighbor. The case $N=3$ is unique, in that the analytical theory shows that the longest relaxation mode is one with all the beads moving in phase. In Fig~\ref{fig2} the mode with the highest amplitude is consistently the one with longest relaxation time.   
    Why is the steady state dynamics largely captured by a normal-mode analysis, and how does the longest lived mode determine the steady state dynamics? 
  The hydrodynamic modes are a key step in constructing the solutions of the dynamical system, which are given for each stretch between consecutive switches by linear superposition of the eigenmodes, and can be obtained analytically for the deterministic (absence of noise) system~\cite{cicuta11w}.
The faster the mode, the more its amplitude has decayed  before each geometric switch.   In other words, it is the amplitude of the longest lived mode that ``dominates'' at the geometric switch, and thus enforces the overall character of the steady state solution.

In the presence of increasing thermal noise, the bead trajectories and mode amplitudes can deviate from the solutions of the deterministic system, as shown in  Fig.~\ref{fig4}, but the solutions remain stable. For $N=7$, like $N=5$ and all other odd $N>3$, there is a degenerate fundamental mode~\cite{cicuta11w}, and the dynamic state shows an alternating amplitude of the projections onto these two modes. The trajectories display a fixed phase relation between beads, and follow each other in the order ${1,3,5,7,2,4,6}$ in Fig.~\ref{fig4}(a,c), and in the equivalent sequence under time reversal in the simulation of Fig.~\ref{fig4}(e).   The nearest neighbors are almost in anti-phase, but delayed by the small interval $T_N/N$. Next-nearest neighbors are almost in-phase, with a delay $2T_N/N$, and so on, describing a propagating wave. Depending on the initial conditions, at low noise the system will fall into a state where  intervals are either positive or  negative going around the ring. This dynamical state  has a propagating phase.   At higher noise (i.e. higher temperature, or weaker coupling), whilst the system remains overall synchronized, the propagating character is lost over long times because the  system is able to flip between the two equivalent states.



The simple actively driven oscillator experiments presented here highlight the importance of geometry in determining the leading properties of the collective steady state. One may envision that in  biological systems, which present complex disordered arrangements, and a vast number of oscillators, the behavior highlighted by these simple arrays could represent the local dynamics in tightly coupled sub-systems. In this perspective, the small-system patterns of behavior can be thought of ``dynamical motifs'', linked simply to the geometrical arrangement of beating elements, and that can be analyzed to infer the properties of the individual oscillators.

\textbf{Acknowledgement} We acknowledge useful discussions with M.Leoni, N.Bruot, E.Onofri, B.Bassetti and M.Baraldi. This work was supported by the Marie  Curie Training Network ITN-COMPLOIDS (FP7-PEOPLE-ITN-2008 No.~234810).

\bibliography{BIBDATAv40}

\clearpage
\newpage

\textbf{Supplementary Material}

\begin{figure}[h!]
\begin{center}
\epsfig{file=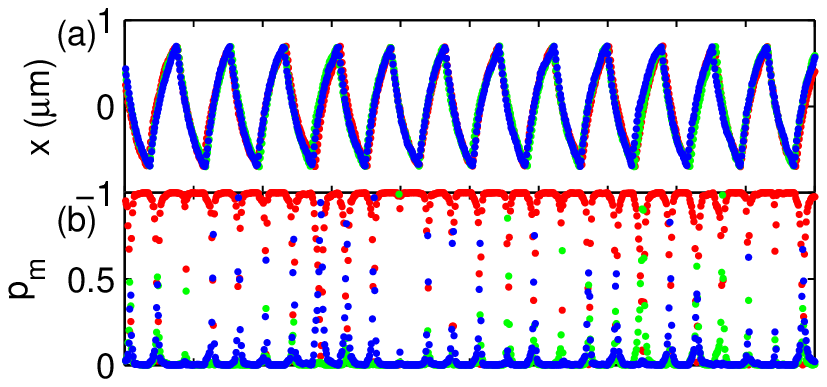, width=7cm}\\
\epsfig{file=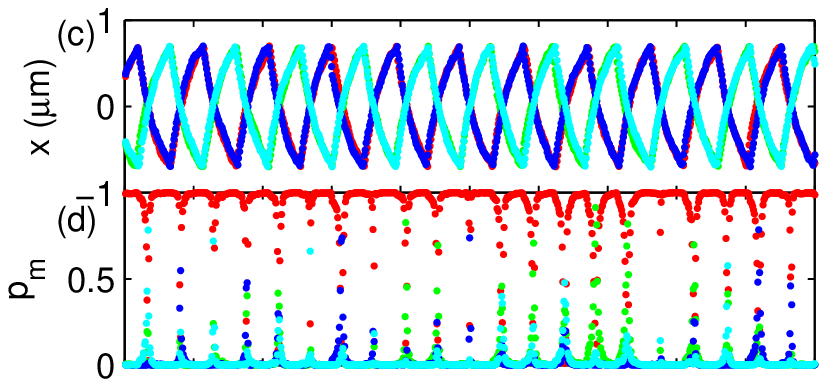, width=7cm}\\
\epsfig{file=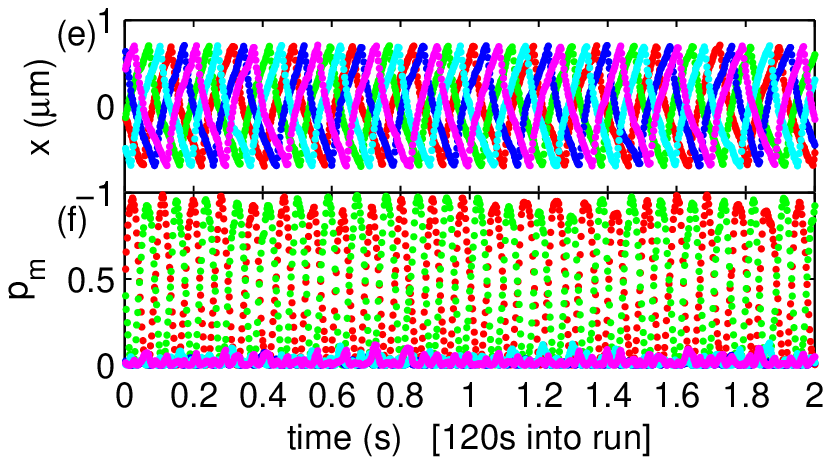, width=7cm}\\
\caption{Supplementary Figure 1 \textbf{Brownian Dynamics simulations with Oseen coupling match the experimental results. } With parameters
and markers matching those of the  experiments of Figure~\ref{fig2} in the Letter. These displacements obtained computationally show that
the description of hydrodynamic coupling via Oseen's tensor is valid.   }
\end{center}
\end{figure}


\end{document}